\documentclass[showpacs,twocolumn,aps,prl,floatfix,superscriptaddress,noshowpacs]{revtex4}
\usepackage{amsthm} 

\usepackage{amsmath,amssymb,graphicx,bm,color,mathrsfs,verbatim,epstopdf,dcolumn,dsfont,hyperref}

\def\beq{\begin{equation}}
\def\eeq{\end{equation}}
\def\bea{\begin{eqnarray}}
\def\eea{\end{eqnarray}}

\usepackage{epstopdf}

\begin{document}

\title{Prethermal Floquet Steady States and Instabilities in the Periodically Driven, Weakly Interacting Bose-Hubbard Model}%

\author{Marin Bukov}
\email{mbukov@bu.edu}
\affiliation{Department of Physics, Boston University, 590 Commonwealth Ave., Boston, MA 02215, USA}

\author{Sarang Gopalakrishnan}
\affiliation{Department of Physics, Harvard University, 17 Oxford Street, Cambridge, MA 02138, USA}

\author{Michael Knap}
\affiliation{Department of Physics, Harvard University, 17 Oxford Street, Cambridge, MA 02138, USA}
\affiliation{Department of Physics, Walter Schottky Institute, and Institute for Advanced Study, Technical University Munich, 85748 Garching, Germany}

\author{Eugene Demler}
\affiliation{Department of Physics, Harvard University, 17 Oxford Street, Cambridge, MA 02138, USA}

\date{\today}

\begin{abstract}
We explore prethermal Floquet steady states and instabilities of the weakly interacting two-dimensional Bose-Hubbard model subject to periodic driving. We develop a description of the nonequilibrium dynamics, at arbitrary drive strength and frequency, using a weak-coupling conserving approximation. We establish the regimes in which conventional (zero-momentum) and unconventional [$(\pi,\pi)$-momentum] condensates are stable on intermediate time scales. We find that condensate stability is \emph{enhanced} by increasing the drive strength, because this decreases the bandwidth of quasiparticle excitations and thus impedes resonant absorption and heating. Our results are directly relevant to a number of current experiments with ultracold bosons.
\end{abstract}



\maketitle

Periodically driven systems\cite{shirley_65,sambe_73,breuer_90,breuer_91} often exhibit exotic phenomena that are absent in their non-driven counterparts\cite{goldman_14, bukov_14,eckardt_15}. Classic examples include the Kapitza pendulum and the periodically kicked rotor. Recently, periodically modulating optical lattices has attracted interest as a way of controlling hopping processes\cite{dunlap_86,eckardt_05,lignier_07,zenesini_09,creffield_10,struck_11,jotzu_15} in order to engineer gauge fields\cite{jaksch_05,mueller_04,struck_13,miyake_13,aidelsburger_13,atala_14,kennedy_15,flaeschner_15}, topological band structures\cite{oka_09,kitagawa_11,jotzu_14,aidelsburger_14,lindner_11,rudner_13,bukov_15_SW}, and associated exotic states of matter. 
Such exotic states are known to exist in noninteracting systems and in certain mean-field models; the extent to which they survive in the presence of interactions is a central open question.
It is believed, from the eigenstate thermalization hypothesis\cite{deutsch_91,srednicki_94,rigol_08}, that driven interacting systems will generically heat up to infinite temperature at sufficiently late times\cite{dalessio_13,dalessio_14,lazarides_14,lazarides_14_2,ponte_15,abanin_15,kuwahara_15,mori_15}. 
Nevertheless, in some parameter regimes these heating times will be parametrically slower than the system's characteristic time scales. In that case, the system will rapidly approach a ``prethermalized'' Floquet steady state\cite{berges_04_prl,canovi_15,kuwahara_15,abanin_15_2}, which governs the dynamics until the much later heating time scales.

In the present work, we study these prethermal states in the weakly interacting, two-dimensional, periodically-driven Bose-Hubbard model (BHM). The regime we explore is directly relevant to experiments\cite{lignier_07,zenesini_09,struck_11,struck_13,aidelsburger_13,miyake_13,aidelsburger_14,kennedy_15,flaeschner_15}, in which weak interactions are present. We employ a self-consistent weak-coupling conserving approximation (WCCA) which treats the coupled nonlinear dynamics of the condensate and the quasiparticle spectrum while neglecting collisions between quasiparticles. This approximation is justified at weak coupling since nonlinearities are important at much shorter times than the collisional time scales. 

\begin{figure}[b]
	\centering
	\includegraphics[width = 1\columnwidth]{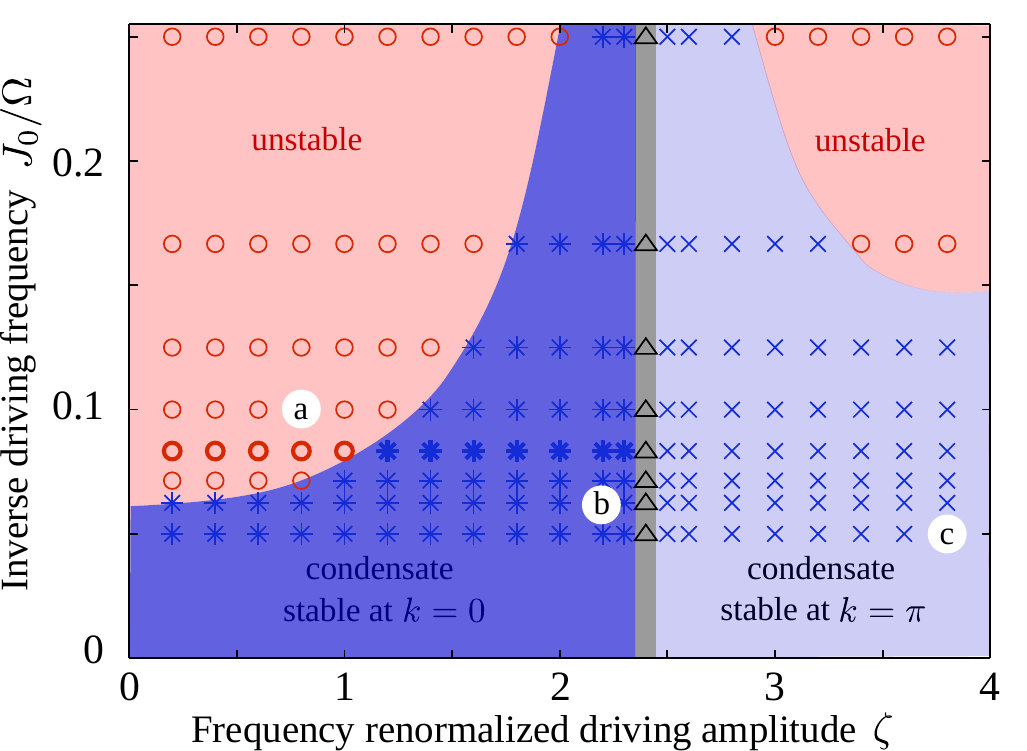}
	\caption{\label{fig:stability_diag} (Color online) Stability diagram of the driven BHM for $U/J_0=0.2$. In the pink regions the condensate is unstable as the drive parametrically excites pairs of quasiparticles. In contrast, in the blue regions the condensate is stable on intermediate time scales. In the grey shaded region around $\zeta \approx 2.405$ the system is strongly correlated (see text). The symbols represent numerical WCCA results; the boundaries are given by the analytical expression Eq.~\eqref{phasebound}. Points marked (a), (b), (c) correspond to the panels in Fig.~\ref{fig:n_k}. }
 
\end{figure}

Within the WCCA, we find a phase diagram (Fig.~\ref{fig:stability_diag}) featuring at low drive frequency a regime in which the superfluid state is \emph{already} unstable within Bogoliubov theory, owing to the resonant creation of quasiparticle pairs, and a regime (at high drive frequency) where the superfluid is stable. In the WCCA, there is a sharp phase transition between these; when effects beyond weak coupling are included, there will be a qualitative difference in heating rates. Thus, in the ``stable'' regions of Fig.~\ref{fig:stability_diag}, the system initially reaches a prethermalized superfluid state---featuring a nonequilibrium quasiparticle distribution---and then eventually heats up. For strong driving, the prethermalized superfluid 
state is exotic, involving condensation at momentum $\boldsymbol{\pi} = (\pi, \pi)$. The existence of this exotic phase in the high-frequency limit has previously been established\cite{dunlap_86,eckardt_05,zenesini_09}; we find that it persists for intermediate frequencies as well. 

Remarkably, we find that the stable phase is \emph{enhanced} for intermediate drive strengths, since
the drive both creates quasiparticle pairs when this is a resonant process, and decreases the effective hopping rate and thus the effective bandwidth of quasiparticle excitations. 
A key conclusion of our work is that, for weak interactions but general drive amplitude and frequency, the condensate becomes unstable when the drive frequency is parametrically resonant with the \emph{drive-renormalized} time-averaged bandwidth. Therefore, parametric resonance occurs at \emph{lower} frequencies when the drive strength is ramped up.

\emph{Model.---}We consider the Bose-Hubbard model on a square lattice in the presence of a circularly-polarised time-periodic force ${\bf E}(t) = A \left(\cos\Omega t,\sin\Omega t\right)^T$:
\begin{equation}
H_\text{lab}(t) \! = \! -\!J_0\!\sum_{\langle ij\rangle }\! b^\dagger_i b_j \! + \!\! \sum_j \! \left[ \frac{U}{2} n_j( n_j\!-\!1) \!\!+ \! {\bf E}(t)\!\cdot\!{\bf r}_j  n_j\!\right].
\label{eq:Hlab(t)} 
\end{equation}
The operator $ b^\dagger_j$ creates a boson on lattice site ${\bf r}_j$. The tunnelling and interaction strength are denoted by $J_0$ and $U$, respectively. To achieve non-trivial dynamics in the high-frequency regime, we scale the driving amplitude linearly with the driving frequency $A\sim\Omega$.\cite{bukov_14}; we define $\zeta \equiv A/\Omega$. We transform this Hamiltonian into a rotating frame (cf.~supplementary material\cite{supp}), giving:
\begin{equation}
H(t) \! =\! -\!J_0\!\sum_{\langle ij\rangle} \! e^{i{\bf \mathcal{A}}(t)\cdot\left({\bf r}_i - {\bf r}_j\right)} b^\dagger_i b_j +\frac{U}{2}\sum_j n_j( n_j-1).
\label{eq:Hrot(t)}
\end{equation}
Thus, in the rotating frame, the system experiences an effective time-dependent gauge potential ${\bf \mathcal{A}}(t) = \zeta \left(\sin\Omega t,-\cos\Omega t\right)^T$. The time evolution of $U(1)$-invariant quantities (and thus the stability) remains the same in both frames\cite{bukov_14_pra}.

\emph{Method.} To study the driven system at arbitrary frequencies, we employ a self-consistent, weak-coupling conserving approximation (WCCA). The WCCA involves deriving equations of motion from a two-particle irreducible effective action\cite{cornwall_74} within the nonequilibrium Schwinger-Keldysh formalism\cite{bauer_14,babadi_15}, keeping only diagrams to first order in $U$ (see~\cite{supp}). 
Unlike simple perturbation theory or Bogoliubov theory, the WCCA respects unitarity and conservation laws\cite{hohenberg_65}, and thus gives physically sensible results for all times; in particular, it allows the exponential growth of unstable modes to be cut off by the resulting depletion of the condensate. While the WCCA is \emph{not} guaranteed to yield a gapless excitation spectrum\cite{hohenberg_65, griffin_95}, the low-frequency behavior of the spectrum is irrelevant for the phenomena discussed here. 
Our approach is equivalent to a fully self-consistent, time-dependent Hartree-Fock-Bogoliubov (HFB) approximation\cite{griffin_95,rey_04}; our formulation, however, can more readily be extended to higher orders in $U$.

\begin{figure}[t]
	\includegraphics[width=\columnwidth]{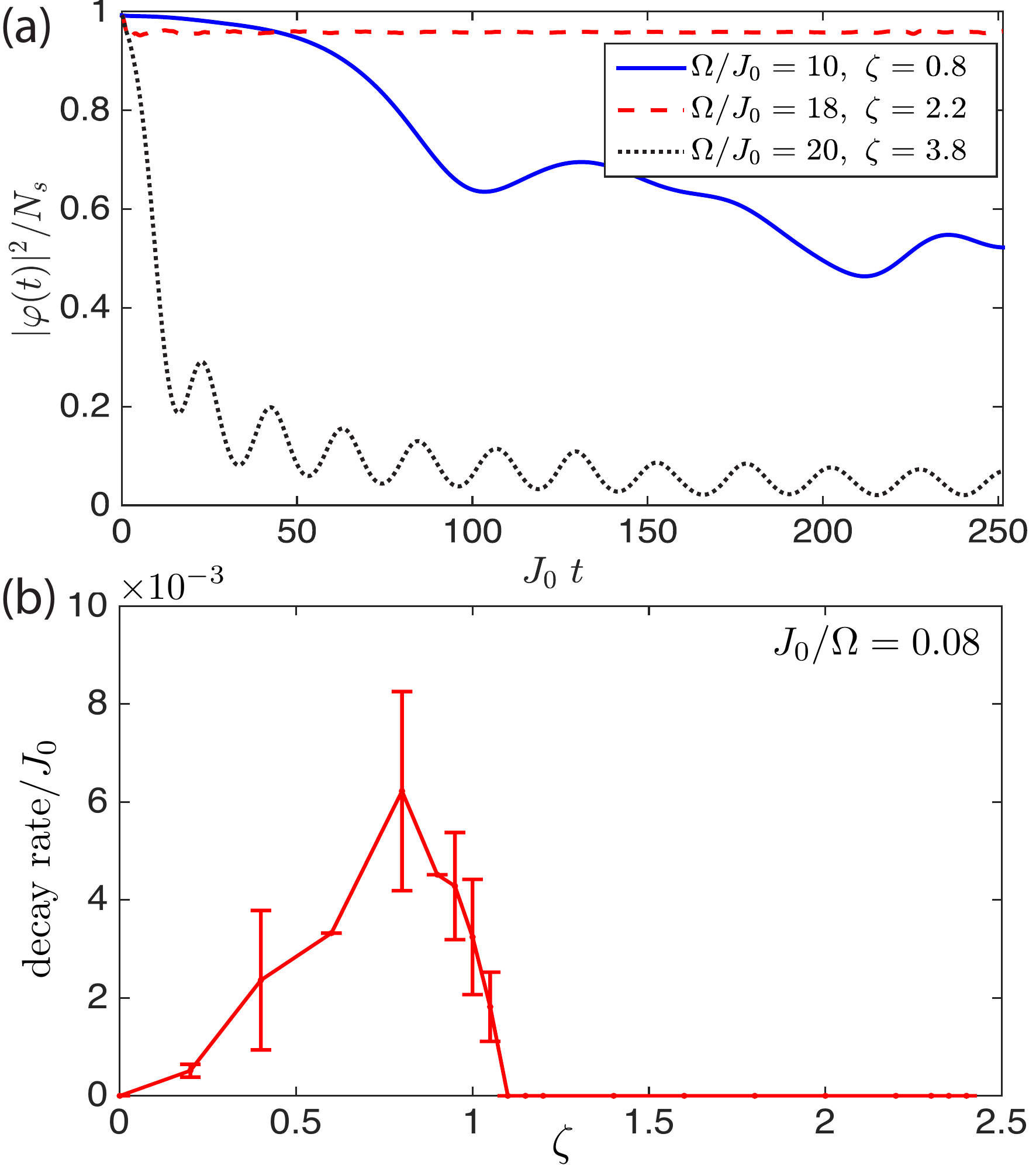}
	\caption{\label{fig:condensate}(Color online). (a) Time evolution of the condensate fraction for $801$ driving cycles, starting from a Bogoliubov initial state localised at ${\bf k} = \mathbf{0}$ for $U/J_0=0.2$. (b) Decay rate to $75\%$ of the condensate curves for $\Omega/J_0 = 12$ (boldface points in Fig.~\ref{fig:stability_diag}). Error bars are set by the difference of the inverse times, determined by the first and last time the curve passes through $3/4$ taking into account the oscillatory behaviour.  }
\end{figure}

The WCCA equations of motion\cite{supp} were solved numerically.
For the results presented here, we prepared the system on a $N_s = 100\times 100$ lattice in the ground state of Bogoliubov theory. We allow for a macroscopic population of the ${\bf k}={\boldsymbol\pi}$ mode to allow for a condensate at momentum $\boldsymbol{\pi}$. To study the nonequilibrium dynamics, we abruptly turn on the periodic drive and propagate the initial state for $801$ driving cycles using Eqs.~(15) and (16) of\cite{supp}. We checked that the results are insensitive to system size.

\emph{Stability diagram}.---The stability phase diagram is shown in Fig.~\ref{fig:stability_diag}. 
Previous work has investigated the driven Bose-Hubbard model\cite{buchleitner_03,kolovsky_03,tomadin_07,kolovsky_09,creffield_09,kolovsky_11,parra-murillo_13,parra-murillo_15} and related models\cite{zhang_04, choundhury_15, bilitewski_15,choundhury_15_2,stoeferle_04,schori_04,kraemer_05,tozzo_05,carusotto_09} using various approximation schemes; we go beyond these works by treating both the condensate and quasiparticle sectors, including the feedback between them. Thus, we are able to explore instabilities originating in either sector on equal footing.

We first discuss two analytically tractable limits, corresponding to high-frequency driving (i.e., going along the $x$ axis of Fig.~\ref{fig:stability_diag}) and to low-amplitude driving (i.e., going along the $y$ axis). In the first case, the dynamics is approximately governed by an effective \emph{time-average} Hamiltonian\cite{goldman_14,bukov_14}:
\begin{equation}
H_\text{ave} = -J_\text{ave}(\zeta)\sum_{\langle ij\rangle } b^\dagger_i b_j + \frac{U}{2}\sum_j  n_j( n_j-1).
\label{eq:Heff}
\end{equation}
The periodic modulation renormalizes the hopping to $J_\text{ave}(\zeta) = J_0\mathcal{J}_0(\zeta)$, where $\mathcal{J}_0(\zeta)$ is the zeroth-order Bessel function of the first kind, which is a damped oscillatory function with the first zero at $\zeta\approx2.4$, the second at $\zeta \approx 5.5$, etc. Thus, as $\zeta$ is increased, the time-averaged hopping decreases, until the dispersion flattens at $\zeta \approx 2.4$. For $\zeta > 2.4$ the dispersion flips sign, and acquires a stable minimum at $\boldsymbol{\pi} = (\pi, \pi)$. Thus, in the high-frequency limit the condensate at ${\bm 0}=(0,0)$ is stable when $\zeta < 2.4$, whereas the condensate at $\boldsymbol{\pi}$ is stable when $2.4\lesssim \zeta \lesssim 5.5$. Moreover, for commensurate filling, the superfluid phase should transition into a Mott insulating state around $\zeta = 2.4$ determined by the phase boundary 
$J_\text{ave}(\zeta)/U \lesssim 0.06$.\cite{capogrosso-sansone_08,knap_10} This transition regime, marked by the thin vertical strip in Fig.~\ref{fig:stability_diag}, is beyond the validity of the WCCA; our WCCA simulations in this regime give oscillatory behavior, see~\cite{supp}.

A second analytically tractable limit is that of weak driving, at arbitrary $\Omega$. The dominant effects can be inferred from linear stability analysis around the non-driven state. In terms of Bogoliubov quasiparticle operators $\gamma_{\mathbf{k}}$, the system-drive coupling includes terms of the form $e^{i \Omega t} \gamma^\dagger_{\mathbf{k}} \gamma^\dagger_{\mathbf{-k}}$, involving the emission of pairs of quasiparticles from the condensate.
The emission rate is related to the density of states of two-quasiparticle excitations at $\Omega$. Specifically, if the non-driven system has quasiparticle excitations at energy $E_{\mathbf{k}}, E_{\mathbf{-k}}$ such that $\Omega = E_{\mathbf{k}} + E_{\mathbf{-k}}$, absorption will occur and the system will be unstable. On the other hand, if $\Omega \geq 2 W$, where $W \approx 2zJ_0$ is the approximate bandwidth of Bogoliubov excitations, then absorption does not occur and the system is stable. 

\begin{figure*}
	\begin{minipage}{\textwidth}
		\includegraphics[width=\columnwidth]{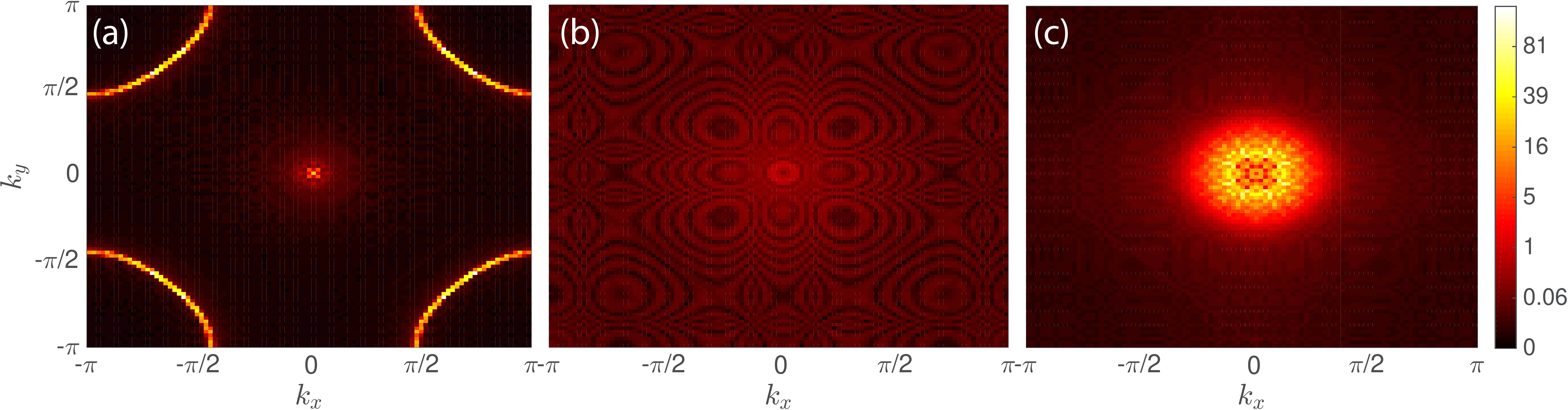}
		\caption{\label{fig:n_k}(Color online). Snapshot of the momentum distribution $n_{\bf k} = \langle b^\dagger_{\bf k} b_{\bf k}\rangle - n_0\delta_{\bf k,0}$ after $801$ driving cycles starting from a Bogoliubov initial state localised at ${\bf k} = {\bf 0}$ for $U/J_0=0.2$. Panel (a) is in the unstable regime where the condensate is depleted due to parametric resonance. The bosons are excited by the drive to the quasienergy surface $\Omega = 2E_\text{ave}(\mathbf{k})$ (bright yellow-white circle around ${\bf k} =\boldsymbol{\pi}$) where they occupy sharp peaks (white pixels). Panel (b) is in the regime where the condensate is stable on the pre-thermal time scales. In panel (c), the system is dynamically unstable due to the dispersion being inverted. The bright disc of excitations around ${\bf k} = {\bf 0}$ corresponds to dynamically unstable modes. The parameters are (a) $\Omega/J_0=10$, $\zeta = 0.8$, (b) $\Omega/J_0=18$, $\zeta = 2.2$, and (c) $\Omega/J_0=20$, $\zeta = 3.8$. }
	\end{minipage}
\end{figure*}

Combining the insights from these two limits allows us to understand the entire stability phase diagram. The drive creates pairs of \emph{renormalized} Bogoliubov quasiparticles, which have an effective bandwidth $W_{\text{ave}} \approx 2z J_{\text{ave}}(\zeta)$. 
We define $W_\text{ave} \equiv \text{max}_k[E_\text{ave}({\bf k})] - \text{min}_k[E_\text{ave}({\bf k})]$ as the time-averaged Floquet-Bogoliubov bandwidth; in terms of this, the stability condition reads
\begin{eqnarray}
\Omega_c > 2W_\text{ave} \Leftrightarrow \text{stable}.
\label{eq:stability_bdary}
\end{eqnarray}
Equation~\eqref{eq:stability_bdary} is consistent with our numerical results (Fig.~\ref{fig:stability_diag}). This result is unexpected---since the time-averaged Hamiltonian is valid at infinite frequency whereas parametric resonance is a low-frequency phenomenon---
but can be understood as follows. The hopping matrix element in the driven system can be expanded as $J(t) \sim J_0 \sum\nolimits_n \mathcal{J}_n(\zeta) \exp(i n \Omega t)$. We absorb the time-independent $n = 0$ component in the unperturbed Hamiltonian, and treat the $n = 1$ term, which oscillates at $\Omega$, perturbatively. The perturbation is small for $U \ll \Omega$, because the matrix element for creating two quasiparticles is proportional to both $\mathcal{J}_1(\zeta)$ [which need not be small] and $U$ [which is assumed to be small]. We then use parametric instability analysis\cite{supp} with the renormalized dispersion, and conclude that an instability occurs when $\Omega = 2 W_{\text{ave}}$. When $\Omega/J_0\gg 1$, the critical driving frequency is given by
\begin{equation}\label{phasebound}
\Omega_c(\zeta) = 4\sqrt{zJ_\text{ave}(\zeta)(zJ_\text{ave}(\zeta)+n_0 U)}.
\end{equation}
Note that in the present case, resonant absorption occurs for drive strengths up to \emph{twice} the single-particle bandwidth; by contrast, in noninteracting systems, no absorption occurs for $\Omega > W_\text{ave}$. The presence of absorption at frequencies exceeding the single-particle bandwidth is generic in interacting systems.

\emph{Condensate evolution}.---Figure~\ref{fig:condensate} [panel (a)] shows the evolution of the condensate fraction in various regimes: in the parametrically unstable regime (solid blue line), the condensate slowly decays; in the stable regime (dashed red line), it saturates to a prethermalized value, which is generally lower than the Bogoliubov value (since $|J_\text{ave}(\zeta)|<|J_0|$). The system enters a steady-state with constant in time evolution when measured stroboscopically. When the initial condensate is at the band maximum (dash-dotted black line), the condensate decays rapidly. Panel (b) shows the decay rate as a function of drive amplitude in the parametrically unstable regime: note that the decay rate depends not only on drive strength $\zeta$, but also on $U$ and $\Omega$. Very close to the region $\zeta\sim 2.405$ (grey strip in Fig.~\ref{fig:stability_diag}), the WCCA gives strong oscillations of the particle density between the condensates at 
$\boldsymbol{0}$ and $\boldsymbol{\pi}$ (see~\cite{supp}); however, as previously noted, the WCCA is not reliable here. 

A natural further observable is the total energy of the system, which grows in the unstable phases and saturates in the stable phases (see~\cite{supp}).

\emph{(Quasi-)momentum distribution.}---Fig.~\ref{fig:n_k} plots snapshots of the quasimomentum (i.e., lattice momentum) distribution; the time evolution of this quantity is shown in\cite{supp}. Specifically, the quantity plotted is $n_{\bf k} = \langle b^\dagger_{\bf k} b_{\bf k}\rangle - n_0\delta_{\bf k,0}$, i.e., the condensate peak is subtracted. The quasimomentum distribution can be directly accessed through band mapping followed by time-of-flight imaging. Moreover, as we are concerned with a single-band model, one can extract this distribution directly from time-of-flight imaging, by focusing on momenta within the first Brillouin zone.

Figure~\ref{fig:n_k} (a) shows the parametrically unstable case, where quasiparticles are strongly excited around the quasimomentum surface $\{{\bf k}:\Omega = 2E_\text{ave}({\bf k}) \}$ matching the resonance condition. Within Bogoliubov theory, the (time-averaged) excitation intensity should be uniform along this surface. However, as the points along this surface are not symmetry-related, the nonlinearities included in the WCCA favor some points on the excitation surface, as seen in the intensity pattern in Fig.~\ref{fig:n_k} (a). 

Figure~\ref{fig:n_k} (b) shows the \emph{stable} case. Here, by contrast with panel (a), the quasiparticle population remains low throughout the Brillouin zone. As expected from Bogoliubov theory, bosonic modes satisfying $J_{\text{ave}}(\mathbf{k}) \alt U$ should have appreciable occupation in the steady state; this region expands as the dispersion flattens. The intricate patterns in momentum space are due to the abrupt turn-on of the drive---which initializes the Floquet-Bogoliubov quasiparticle states out of equilibrium---and are absent when the drive is instead gradually ramped up. These patterns evolve nontrivially with time (see~\cite{supp}). 

Finally, Fig.~\ref{fig:n_k} (c) illustrates the case in which the initial state is a condensate at ${\bf k} = {\bf 0}$, but the dispersion is inverted ($\zeta > 2.4$) so that the only \emph{stable} condensate is supported at ${\bf k} = \boldsymbol{\pi}$. Thus the initial state is unstable regardless of $\Omega$. Let us consider the infinite-frequency limit; which amounts to a sudden quench of the single-particle dispersion. Computing the Bogoliubov spectrum around a condensate at ${\bf k} = {\bf 0}$ in an inverted dispersion, we find that modes with momenta near ${\bf k} = {\bf 0}$ acquire imaginary frequencies (and thus grow exponentially), whereas modes with large momenta are \emph{stable}~\footnote{This might seem counterintuitive, as the larger-momentum modes have ``more negative'' energies; note, however, that in the $U \rightarrow 0$ limit, \emph{all} modes are stable as there are no decay processes.}. The 
unstable 
modes are determined by the condition $\varepsilon_\text{ave}({\bf k})+zJ_0 < 2n_0 U$, where $\varepsilon_\text{ave}({\bf k})$ is the single-particle Floquet dispersion~\eqref{eq:Heff}. These modes are dynamically stabilized due to the nonlinear feedback of the self-consistent treatment\cite{babadi_15}. 
Our numerical results with the WCCA confirm this picture: the unstable modes at small quasimomenta acquire large populations, whereas the large-quasimomentum modes do not. 
This behavior is \emph{specific} to the WCCA; in a real system it will correspond to intermediate-time dynamics $t\lesssim J_0/U^2$. On longer times, collisions between quasiparticles should cause large occupation numbers across the Brillouin zone, see~\cite{supp}.

\emph{Discussion}.---We briefly outline the validity of the WCCA in the three regimes of interest (for details see~\cite{supp}). In the parametrically unstable regime, the stability analysis suggests that unstable modes grow at the rate $\Gamma \sim U n_0 J_0\mathcal{J}_1(\zeta)/W_\mathrm{ave}$, while the momentum arcs in Fig.~\ref{fig:condensate} (a) decay at a Golden Rule rate $\sim U^2 n_0 n_{\mathbf{k}}/W_\text{ave}$. Hence, as long as $U n_{\mathbf{k}} < J_0\mathcal{J}_1(\zeta)$, the formation rate is greater than the decay rate and the WCCA is reliable. In the stable region, the condensate fraction $n_0$ remains large, and the WCCA remains valid, until very late times, when resonant absorption involving $m=\Omega/W_\text{ave}$ quasiparticles becomes dominant. For large $\Omega$, this is a very high-order and therefore very slow process. Finally, in the dynamically unstable phase, the WCCA physics is valid up to times $W_\text{ave}/U^2$ (the collisional time scale). Thus, at weak coupling, there is a parametrically large window between $1/\sqrt{J_\mathrm{ave}(\zeta)U}$ and $W_\text{ave}/U^2$ where the WCCA description is correct.  

The main experimental prediction of this work---a parametric change in heating rates as a function of drive amplitude and frequency---can be measured in present-day experiments, which are naturally in the weak-coupling regime. For the experiment in Ref.~\onlinecite{aidelsburger_14} the parameters were chosen as $U/J_0 \approx 0.1$, $\Omega/J_0\approx 20$, and $\zeta \approx 0.6$, which is within the regime we considered. For realistic experiments in optical lattices, the presence of higher bands can lead to instability even at high drive frequencies $\Omega$. In this case there are three regimes: (i) if $\Omega$ is less than twice the renormalized bandwidth $W_\text{ave}$ of the lower band, the system is parametrically unstable as discussed above; (ii) if $\Omega$ is larger than $2 W_\text{ave}$, smaller than the band gap to the upper band, and furthermore chosen such that any $n$-photon resonances to higher bands~\cite{weinberg_15} are suppressed, then the system is stable within WCCA. (iii) if $\Omega$ exceeds the band gap, the drive can mediate interband transitions, leading to instability again. For a square optical lattice with typical lattice potential $V_\text{latt} = 10E_\text{recoil}$, $E_\text{recoil} = h\times 4\; \text{kHz}$, the bandwidth of the lowest band is $W_0 = 4J_0 = h\times 0.3\; \text{kHz}$ [the time-averaged bandwidth $W_\text{ave}$ is reduced by a factor of $\mathcal{J}_0(\zeta)$], and the gap to the second Bloch band is $\Delta = 4.57E_\text{recoil} = h\times 18.28\; \text{kHz}$.

Although we focused on a square lattice, the arguments generalize to other lattices, such as the honeycomb lattice, in which topologically non-trivial states exist. Note that topological gaps in \emph{mechanically shaken} optical lattices scale as $\Omega^{-1}$\cite{oka_09,kitagawa_11,jotzu_14}. Hence, in order to engineer topological insulators with large gaps (and a large region of non-zero Berry 
curvature around them), it is desirable to go to lower frequencies. Our results impose a fundamental limit for \emph{weakly-interacting bosonic} systems on how small the frequency can be, since for $\Omega < 2W_\text{ave}$ the system becomes unstable. More generally, our results suggest that conserving approximations, whether controlled by weak coupling or some other parameter as in large-$N$ models\cite{mosche_03,sciolla_13,smacchia_15,babadi_15,chandran_15}, are ways of exploring dynamical phase transitions in models that are both \emph{interacting} (unlike free-particle models) and \emph{finite-dimensional} (unlike the Kapitza pendulum). The critical properties of such 
transitions are a fruitful theme for future work. Although in practice such phase transitions will be smeared out by higher-order effects, the associated crossovers should still be experimentally observable.

\emph{Acknowledgements}. We thank D.~Abanin, M.~Babadi, I.~Bloch, L.~D'Alessio, E.~Dalla Torre, N.~Goldman, M.~Kolodrubetz, A.~Polkovnikov, and U.~Schneider for interesting and fruitful discussions, and especially acknowledge the help of M.~Lohse in determining the experimentally relevant parameters.
The authors acknowledge support from the NSF grant DMR-1308435, Harvard-MIT CUA, AFOSR New Quantum Phases of Matter MURI, the ARO-MURI on Atomtronics, ARO MURI Quism program, AFOSR FA9550-13-1-0039 and BSF 2010318, and Technical University Munich - Institute for Advanced Study, funded by the German Excellence Initiative and the European Union FP7 under grant agreement 291763.

\begin{widetext}

\section{Transformation to the Rotating Frame and Stability Analysis  }

In this appendix, we begin by discussing the transformation of the driven BHM from the lab frame to the rotating frame, before doing a Bogoliubov stability analysis. The rotating frame is defined by the unitary transformation
\begin{eqnarray}
V(t) = \exp\left(-i\left[ \int_0^t{\bf E}(t')\mathrm{d}t' + \mathcal{A}(0) \right]\cdot \sum_j {\bf r}_j n_j \right),\ \ \ \ \
|\psi_\text{rot}(t)\rangle = V(t)|\psi_\text{lab}(t)\rangle,\ \ \ \ \
b_j\to b_j e^{-i\mathcal{A}(t)\cdot{\bf r}_j},
\end{eqnarray}
where ${\bf \mathcal{A}}(t) = \zeta \left(\sin\Omega t,-\cos\Omega t\right)^T$. This time-dependent change of basis is equivalent to the standard gauge transformation in electromagnetism ${\bf E}(t)\sim\partial_t\mathcal{A}(t)$. Physically, the transformation trades the fast time-dependence of the quasimomentum for a time oscillating dispersion relation. The infinite-frequency limit is non-trivial when the amplitude of the gauge potential $\zeta= A/\Omega$ remains finite. This rotation facilitates the analytic computation the time-averaged Floquet Hamiltonian. In fact, going to the rotating frame is equivalent to re-summing an infinite inverse-frequency subseries in the lab-frame\cite{bukov_14}. As a result, the effective hopping matrix element is a non-perturbative function of $\zeta$.

The rotating frame Hamiltonian reads
\begin{eqnarray}
\label{eq:H_rot_frame}
H(t) &=& -J_0\sum_{\langle ij\rangle} \! e^{i{\bf \mathcal{A}}(t)\cdot\left({\bf r}_i - {\bf r}_j\right)} b^\dagger_i b_j +\frac{U}{2}\sum_j n_j( n_j-1)\nonumber\\
&=& H_\text{ave} -\sum_{\langle ij\rangle} \left(J_0 e^{i{\bf \mathcal{A}}(t)\cdot\left({\bf r}_i - {\bf r}_j\right)} - J_\text{ave}\right) b^\dagger_i b_j,
\nonumber
\end{eqnarray}	
where in the second equality we separated out the time average explicitly. The time-average Hamiltonian $H_\text{ave}$ is defined in Eq.~(3) of the main text, and the effective hopping is $J_\text{ave} = J_0\mathcal{J}_0(\zeta)$. To avoid confusion in the notation, we note in passing that this effective Hamiltonian is \emph{not} equivalent to the full Floquet Hamiltonian at any finite driving frequency, see eg.~Refs.~\cite{goldman_14,bukov_14}.

Before we dive into the details of the parametric stability analysis for the driven Bose-Hubbard model, let us demonstrate how to derive the stability criterion with the help of the Rotating Wave Approximation (RWA) [a very similar method was used to study the parametric instability in periodically-driven Luttinger liquids~\cite{bukov_12}]. For this purpose, we choose the parametric oscillator with Hamiltonian $H(t) = \frac{1}{2}\left(p^2 + \omega_0^2x^2 + \alpha\omega_0^2\cos(\Omega t)x^2 \right)$. Writing the Hamiltonian using ladder operators $x = 1/\sqrt{2\omega_0}(\gamma^\dagger + \gamma)$ and $p = i\sqrt{\omega_0/2}(\gamma^\dagger - \gamma)$, and dropping any (time-dependent) constants leads to
\begin{equation*}
H(t) = \omega_0\left(1+\frac{\alpha}{2}\cos\Omega t\right)\gamma^\dagger\gamma + \frac{\alpha\omega_0}{4}\cos\Omega t\left(\gamma^\dagger\gamma^\dagger + \mathrm{h.c.}\right).
\end{equation*}
If we parametrise $\gamma(t) = u'(t) \gamma(t=0) - v'^*(t)\gamma^\dagger(t=0)$, with $u'(t=0) =1$ and $v'(t=0) =0$, we can write Heisenberg's EOM as
\begin{eqnarray}
i\frac{\mathrm{d}}{\rm{d}t}
\left(
\begin{array}{c}
u' \\ v'  
\end{array}
\right)
&=& 
\left(
\begin{array}{cc}
\omega_0 + \frac{\alpha}{2}\omega_0\cos\Omega t & \frac{\alpha}{2}\omega_0\cos\Omega t \\
- \frac{\alpha}{2}\omega_0\cos\Omega t & -\left( \omega_0 + \frac{\alpha}{2}\omega_0\cos\Omega t \right)
\end{array}
\right)
\left(
\begin{array}{c}
u' \\ v'  
\end{array}
\right)
\nonumber\\
&=& \left[ \omega_0\sigma^z + W(t)\right]
\left(
\begin{array}{c}
u' \\ v'  
\end{array}
\right) 
+ \frac{\alpha}{2}\omega_0\cos\Omega t
\left(
\begin{array}{cc}
0 & 1 \\ -1 & 0
\end{array}
\right)
\left(
\begin{array}{c}
u' \\ v'  
\end{array}
\right),
\end{eqnarray}
where the matrix $W(t) = \frac{\alpha}{2}\omega_0\cos(\Omega t)\sigma^z$ has zero time-average and $\sigma^z$ is the Pauli matrix in Bogoliubov space. We now apply the transformation $\tilde u'(t) = e^{i2\omega_0 t}u'(t)$, $\tilde v'(t) = v'(t)$ which brings the EOM into the form
\begin{eqnarray}
i\frac{\mathrm{d}}{\rm{d}t}
\left(
\begin{array}{c}
\tilde u' \\ \tilde v'  
\end{array}
\right)
&=& \left[ \omega_0 + W(t) 
+ \frac{\alpha}{2}\omega_0
\left(
\begin{array}{cc}
0 & e^{-2i\omega_0 t}\cos\Omega t \\ -e^{+2i\omega_0 t}\cos\Omega t & 0
\end{array}
\right)
\right]
\left(
\begin{array}{c}
\tilde u' \\ \tilde v'  
\end{array}
\right).
\end{eqnarray}
So far the treatment of the parametric oscillator EOM has been exact. However, the present form of the equation allows to easily identify the terms responsible for parametric resonance. To this end, we apply the rotating wave approximation (RWA) (i) keeping in mind that the time-average of $W(t)$ vanishes identically, and (ii) dropping any counter-rotating terms. Thus, we find that the dominant contribution to the dynamics appears for $2\omega_0=\Omega_c$, which sets the critical driving frequency on resonance. The resulting effective RWA-EOM assumes the simple form:
\begin{eqnarray}
i\frac{\mathrm{d}}{\rm{d}t}
\left(
\begin{array}{c}
\tilde u' \\ \tilde v'  
\end{array}
\right)
&=&  
\left(
\begin{array}{cc}
\omega_0 & \frac{\omega_0\alpha}{4} \\ -\frac{\omega_0\alpha}{4} & \omega_0
\end{array}
\right)
\left(
\begin{array}{c}
\tilde u' \\ \tilde v'  
\end{array}
\right).
\end{eqnarray}
Diagonalising the matrix on the right-hand side, we find the two Lyapunov exponents $\lambda_{1,2} = \omega_0\pm i\omega_0\alpha/4$. Hence, the maximum instability growth rate is set by $\alpha\omega_0/4$. The stability criterion and the instability growth rate \emph{on resonance} derived above by means of the RWA agree precisely with the standard results obtained using two-time perturbation theory or by other means~\cite{book_LL}.

In the following, we apply the same method and show explicitly the stability analysis for the driven BHM leading to the resonance condition $\Omega_c = 2E_\text{ave}({\bf k})$ of the main text. Had it not been for the subindex $_\mathrm{ave}$, this result follows immediately from the above arguments for the parametric oscillator. We work in Bogoliubov theory which, as we show, captures the onset of instability. We begin by writing the rotating frame Hamiltonian	of Eq.~\eqref{eq:H_rot_frame} in momentum space and apply to it the Bogoliubov approximation. Parametrising the bosonic annihilation operator by $b_{\bf k}(t) = u_{\bf k}(t) b_{\bf k}(t=0) - v^*_{\bf -k}(t)b^\dagger_{\bf -k}(t=0)$, with $u_{\bf k}(t=0) =1$ and $v_{\bf k}(t=0) =0$, the Heisenberg equations of motion read
\begin{eqnarray}
i\frac{\mathrm{d}}{\rm{d}t}
\left(
\begin{array}{c}
u_{\bf k} \\ v_{\bf k}  
\end{array}
\right)
&=& 
\left(
\begin{array}{cc}
\varepsilon({\bf k},t) +zJ_0 +U n_0 & U n_0 \\ -U n_0 & -[\varepsilon({\bf -k},t) +zJ_0 +U n_0]  
\end{array}
\right)
\left(
\begin{array}{c}
u_{\bf k} \\ v_{\bf k}  
\end{array}
\right)\nonumber\\
&=&
\left(
\begin{array}{cc}
\varepsilon_\text{ave}({\bf k}) +zJ_0 +U n_0 & U n_0 \\ -U n_0 & -(\varepsilon_\text{ave}({\bf k}) +zJ_0 +U n_0)  
\end{array}
\right)
\left(
\begin{array}{c}
u_{\bf k} \\ v_{\bf k}  
\end{array}
\right) 
+ 
\left(
\begin{array}{cc}
g_{\bf k}(t) & 0 \\ 0 & -g_{\bf -k}(t)  
\end{array}
\right)
\left(
\begin{array}{c}
u_{\bf k} \\ v_{\bf k}  
\end{array}
\right), 
\end{eqnarray} 
where in the second line we separated the time average. Here $n_0$ is the condensate fraction [of the time-averaged Hamiltonian], and the periodic function $g_{\bf k}(t) = \varepsilon({\bf k},t) - \varepsilon_\text{ave}({\bf k})$. It will prove convenient to first perform a static Bogoliubov transformation $M_{\bf k}(\theta)$, which diagonalises the time-averaged Hamiltonian:
\begin{eqnarray}
\left(
\begin{array}{c}
u_{\bf k} \\ v_{\bf k}  
\end{array}
\right) 
&=& M_{\bf k}(\theta) 
\left(
\begin{array}{c}
u'_{\bf k} \\ v'_{\bf k}  
\end{array}
\right)
= 
\left(
\begin{array}{cc}
\cosh(\theta_{\bf k}) & \sinh(\theta_{\bf k}) \\ \sinh(\theta_{\bf k}) & \cosh(\theta_{\bf k})
\end{array}
\right)
\left(
\begin{array}{c}
u'_{\bf k} \\ v'_{\bf k}  
\end{array}
\right).
\end{eqnarray}
The Bogoliubov angle is defined via $\cosh(2\theta_{\bf k}) = (\varepsilon_\text{ave}({\bf k})+n_0 U)/E_\text{ave}({\bf k})$ and $\sinh(2\theta_{\bf k}) = n_0 U/E_\text{ave}({\bf k})$ with $E_\text{ave}({\bf k})$ the corresponding Bogoliubov dispersion (see main text). The pseudounitary operator $M_{\bf k}(\theta)$ has the property $M^\dagger_{\bf k}(\theta)\sigma^z M_{\bf k}(\theta) = \sigma^z$ [note that $M^{-1}_{\bf k}(\theta)\neq M^\dagger_{\bf k}(\theta)$].

The idea behind performing this Bogoliubov transformation is to bring the time-averaged Hamiltonian in diagonal form. At high-frequencies, the former represents the leading-order Floquet Hamiltonian [when expanded in powers of the inverse frequency], and thus this Bogoliubov transformation brings the state at time $t=0$ to a basis which is close but not equal to the exact Floquet basis [finite $\Omega^{-1}$-corrections are missing]:
\begin{eqnarray}
i\frac{\mathrm{d}}{\rm{d}t}
\left(
\begin{array}{c}
u'_{\bf k} \\ v'_{\bf k}  
\end{array}
\right)
&=&
E_\text{ave}({\bf k})\sigma^z
\left(
\begin{array}{c}
u'_{\bf k} \\ v'_{\bf k}  
\end{array}
\right)
\nonumber\\
&& + 
\left(
\begin{array}{cc}
g_{\bf k}(t)\cosh^2(\theta_{\bf k}) + g_{\bf -k}(t)\sinh^2(\theta_{\bf k})  & 0 \\ 0 &  -g_{\bf -k}(t)\cosh^2(\theta_{\bf k}) - g_{\bf k}(t)\sinh^2(\theta_{\bf k})  
\end{array}
\right)
\left(
\begin{array}{c}
u'_{\bf k} \\ v'_{\bf k}  
\end{array}
\right)
\nonumber\\
&& + \frac{1}{2}\left[ g_{\bf k}(t) + g_{\bf-k}(t) \right]\sinh(2\theta_{\bf k})
\left(
\begin{array}{cc}
0 & 1 \\ -1 & 0  
\end{array}
\right)
\left(
\begin{array}{c}
u'_{\bf k} \\ v'_{\bf k}  
\end{array}
\right).
\end{eqnarray}
Introducing the short-hand notation
\begin{eqnarray}
W_{\bf k}(t) &=&
\left(
\begin{array}{cc}
g_{\bf k}(t)\cosh^2(\theta_{\bf k}) + g_{\bf -k}(t)\sinh^2(\theta_{\bf k}) & 0 \\ 0 & - g_{\bf -k}(t)\cosh^2(\theta_{\bf k}) - g_{\bf k}(t)\sinh^2(\theta_{\bf k})  
\end{array}
\right),
\nonumber\\
h_{\bf k}(t) &=& \frac{1}{2}\left[ g_{\bf k}(t) + g_{\bf-k}(t) \right],
\end{eqnarray}
the EOM readily assumes the form:
\begin{eqnarray}
i\frac{\mathrm{d}}{\rm{d}t}
\left(
\begin{array}{c}
u'_{\bf k} \\ v'_{\bf k}  
\end{array}
\right)
&=&
\left[ E_\text{ave}({\bf k})\sigma^z + W_{\bf k}(t)\right]
\left(
\begin{array}{c}
u'_{\bf k} \\ v'_{\bf k}  
\end{array}
\right)
+ \sinh(2\theta_{\bf k}) h_{\bf k}(t)
\left(
\begin{array}{cc}
0 & 1 \\ -1 & 0
\end{array}
\right)
\left(
\begin{array}{c}
u'_{\bf k} \\ v'_{\bf k}  
\end{array}
\right).
\end{eqnarray}
Notice that the diagonal matrix $W_{\bf k}(t)$ has a zero time average, a property inherited from the function $g_{\bf k}(t)$. In the non-interacting limit, where one can integrate the EOM exactly, we have $\theta_{\bf k}\to 0$ and the time-dependent term $W_{\bf k}(t)$ results in a trivial dynamical phase whose origin can be traced back to the kick operator of Floquet theory\cite{goldman_14,bukov_14}. Thus, we find that at small $U$, the kick operator attains additional contributions due to the Bogoliubov dressing in the diagonal matrix $W_{\bf k}(t)$. Although parametric instability is a phenomenon believed to originate in the micromotion itself [the stroboscopic Floquet Hamiltonian, if it exists as a local operator, is a hermitian operator and therefore has real eigenvalues], in both the weakly-interacting and the non-interacting case this $W_{\bf k}(t)$-term has no significance for the onset of the parametric instability, as we show below. We stress that $W_{\bf k}(t)$ is not the only contribution to the micromotion; part of the latter is due to the function $h_{\bf k}(t)$. Thus, special attention must be paid to the term proportional to $h_{\bf k}(t) \sinh(2\theta_{\bf k})\sim n_0 U h_{\bf k}(t)$. In the instantaneous diagonal basis, this results to the coupling of the form $h_{\bf k}(t)\gamma^\dagger_{\bf -k}\gamma^\dagger_{\bf k}$, and is responsible for the drive-assisted scattering of bosons out of the condensate mentioned in the main text.  

In order to make the nature of parametric resonance visible, we perform yet another unitary time-dependent transformation $\tilde u'_{\bf k}(t) = e^{2iE_\text{ave}({\bf k}) t}u'_{\bf k}(t)$, $\tilde v'_{\bf k}(t) = v'_{\bf k}(t)$. This transformation is needed to bring the relative dynamical phases of $u'_{\bf k}$ and $v'_{\bf k}$ with energy $E_\text{ave}({\bf k})$ at the same footing. In some physical sense, the onset of the parametric instability for a quantum harmonic oscillator is due to an enhanced mismatch of the relative dynamical phases accumulated by the operators $b(t)$ and $b^\dagger(t)$ in the presence of the periodic drive. The EOM now takes the form
\begin{eqnarray}
i\frac{\mathrm{d}}{\rm{d}t}
\left(
\begin{array}{c}
\tilde u'_{\bf k} \\ \tilde v'_{\bf k}  
\end{array}
\right)
&=&
\left[ E_\text{ave}({\bf k}) + W_{\bf k}(t) +
\sinh(2\theta_{\bf k}) 
\left(
\begin{array}{cc}
0 & h_{\bf k}(t)e^{-2iE_\text{ave}({\bf k})t} \\ -h_{\bf k}(t)e^{2iE_\text{ave}({\bf k})t} & 0
\end{array}
\right)
\right]
\left(
\begin{array}{c}
\tilde u'_{\bf k} \\ \tilde v'_{\bf k}  
\end{array}
\right).
\end{eqnarray}

So far the above analysis was rather involved but exact. We can make further progress by Flourier-expanding the periodic function $g_{\bf k}(t) = J_0\sum_{l\neq 0} c_k^l(\zeta)e^{il\Omega t}$ [note that $g_{\bf k}(t)$ has zero mean by definition] which leads to $h_{\bf k}(t) = J_0/2\sum_{l\neq 0} [c_{\bf k}^l(\zeta)+c_{\bf -k}^l(\zeta)]e^{il\Omega t}$. The Fourier coefficients $c_k^l(\zeta)$ are closely related to Bessel functions. Now we apply the rotating wave approximation (RWA) to the above equation. This brings about a pronounced dominant contribution from the drive for $\Omega_c = 2E_\text{ave}({\bf k})$ coming from the slowly rotating oscillatory off-diagonal terms. This behaviour is in precise agreement with the parametric resonance condition from the main text which was verified numerically using the WCCA. Recalling that the time-average of $g_{\bf k}(t)$ vanishes, the dynamics of the system on resonance is governed by the following set of effective equations
\begin{eqnarray}
i\frac{\mathrm{d}}{\rm{d}t}
\left(
\begin{array}{c}
\tilde u'_{\bf k} \\ \tilde v'_{\bf k}  
\end{array}
\right)
&\approx&
\left(
\begin{array}{cc}
E_\text{ave}({\bf k}) & J_0/2\sinh(2\theta_{\bf k}) \left[c^{-1}_{\bf k}(\zeta)+ c^{-1}_{\bf -k}(\zeta)\right] \\ -J_0/2\sinh(2\theta_{\bf k})\left[c^{1}_{\bf k}(\zeta)+ c^{1}_{\bf -k}(\zeta)\right] & E_\text{ave}({\bf k})
\end{array}
\right).
\end{eqnarray}
Diagonalising this matrix, we find the Lyapunov exponents on resonance [we avoid the terminology quasienergies since unlike quasienergies the Lyapunov exponents can be complex numbers], 
\begin{equation}
\epsilon_{1,2} = E_\text{ave}({\bf k})\pm i \frac{J_0}{2}\sinh(2\theta_{\bf k}) \sqrt{ \left[c^{-1}_{\bf k}(\zeta)+ c^{-1}_{\bf -k}(\zeta)\right]\left[c^{+1}_{\bf k}(\zeta)+ c^{+1}_{\bf -k}(\zeta)\right]  },
\end{equation}
whose imaginary part is responsible for the exponential growth of the parametrically-unstable solution within Bogoliubov theory. Notice that $\sinh(2\theta_{\bf k})\sim n_0 U$ defines the instability growth rate which is directly tied to the heating rate at short times in the parametrically unstable regime. At small driving amplitudes, the imaginary part of the Lyapunov exponent scales linearly with the drive strength $\zeta=A/\Omega$, as expected. Higher-order photon absorption resonances can be taken into account by using the higher-order Fourier coefficients $c^{\pm l}_{\bf k}(\zeta)$. The behaviour of the system in the vicinity of the parametric resonance, on the other hand, can be analysed by introducing a small detuning $\delta = \Omega-2E_{\bf k}$. 

Finally, we remark that, the above analysis within Bogoliubov theory is not expected to produce the correct dynamics in the unstable regimes due to the lack of particle number conservation. Instead, one needs to further develop the perturbation theory by extending it to the WCCA.

\begin{figure}[h!]
	\centering
	\includegraphics[width = 0.5\columnwidth]{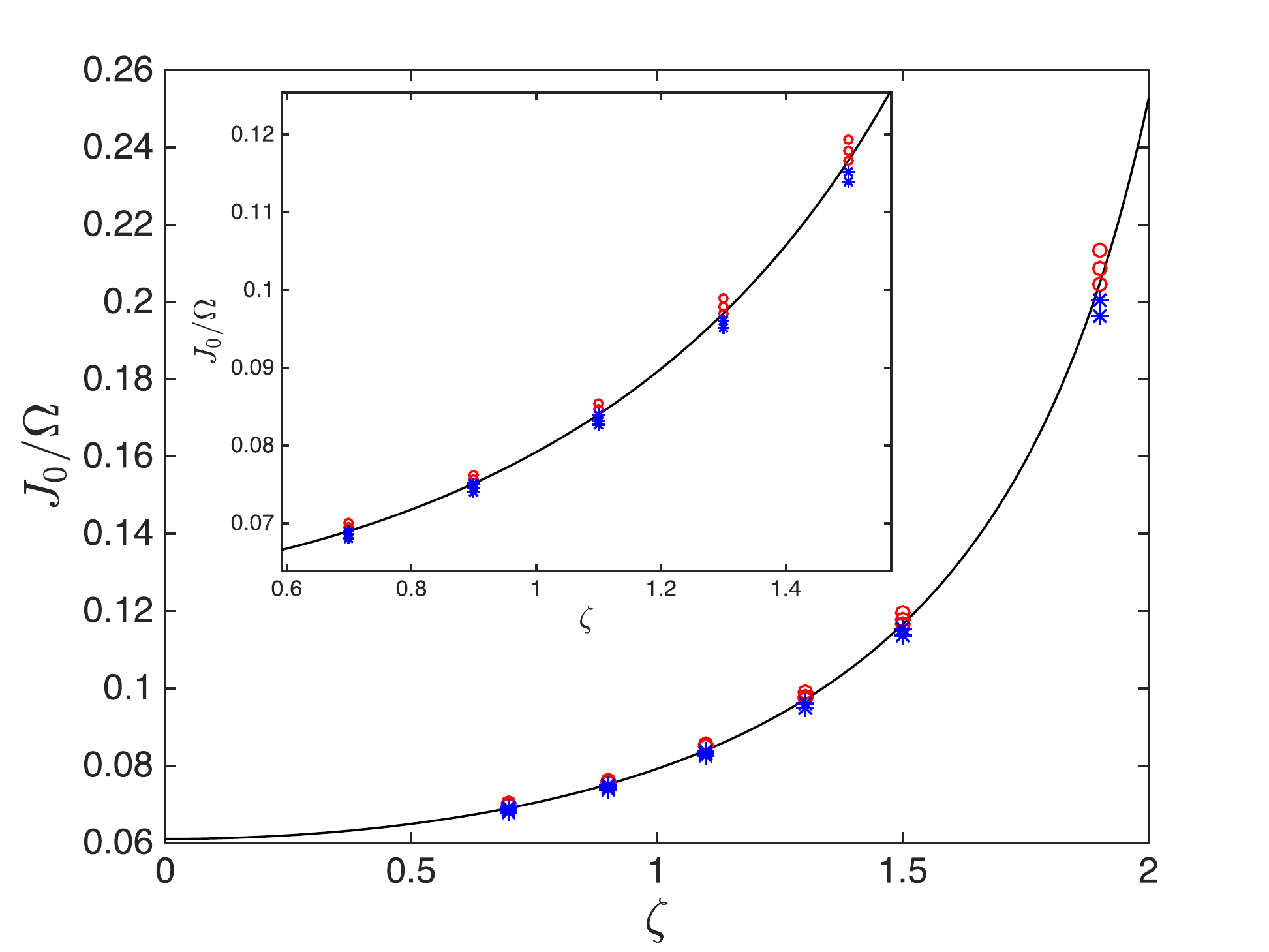}
	\caption{\label{fig:ph_diag_resolved} Comparison between the theoretical Bogoliubov stability boundary (black solid line) and the numerical solution to the WCCA equations discussed in the main text. Similarly to Fig.~1 of the main text, red circles mark the unstable while blue stars - the stable phase. The inset zooms in the small-$\zeta$ region to provide a better resolution. }  
\end{figure}

Figure~\ref{fig:ph_diag_resolved} shows a direct comparison between the stability criterion for the transition boundary $\Omega>2E_\mathrm{ave}({\bf k})\Leftrightarrow \text{``stable"}$, shown as a solid black line, and the numerical solution to the WCCA equations. Numerically, the condensate is defined to be stable whenever, after $800$ periods of stroboscopic evolution, no decay is visible. In general, we find an excellent agreement between the two approaches; the only difference comes for the points lying right on the transition boundary which appear to be stable for small $\zeta$ and unstable for large $\zeta$. We attribute this to the uncertainty in determining the numerical phase boundary: indeed, coming from the stable phase the lifetime of the condensate can be very long, thus exceeding the $800$ periods of evolution time.

\section{Derivation of the equations of motion within the Weak-Coupling Conserving Approximation}

In this appendix, we derive the equations of motion (EOM) using the weak-coupling conserving approximation (WCCA). We are interested in studying the periodically driven Bose-Hubbard model on a 2D lattice:
\begin{equation}
H(t) = -\sum_{ij}J_{ij}(t) b^\dagger_i b_j + \text{h.c.} + \frac{U}{2}\sum_j  n_j( n_j-1),
\end{equation} 
In order to treat the spontaneous symmetry breaking of the condensate efficiently, we introduce the Bogoliubov spinor for the bosonic fields $ b\to b_{a}$, with $a = 1,2$, where $ b_{1} = b$ and $b_{2} = b^*$. Adopting the notation $(j,t) = x$, the time-dependent action can be cast into the compact form
\begin{eqnarray}
S[ b, b^*]  &=& S_0 + S_\text{int}\nonumber\\
S_0[ b,b^*]  &=& \frac{1}{2}\int_C\mathrm{d}x  b^*_{a}(x)\left(G_\text{free}^{(-1)}\right)_{ab}(x,y) b_{b}(y)\nonumber\\
S_\text{int}[b,b^*]  &=& -\frac{U}{2}\int_C\mathrm{d}x\mathrm{d}y \delta_C(x-y)  b^*(x) b^*(x) b(x) b(x).
\end{eqnarray}  
where the integral over time is taken along the Keldysh roundtrip contour ${C}$\cite{kamenev_04,berges_04,cornwall_74} and we introduced the delta function $\delta_C(x-x') = \delta_C(t-t')\delta_{jj'}$. In Bogoliubov space, the noninteracting Green's function has thus the form
\begin{eqnarray}
\left(G^{(-1)}_\text{free}\right)_{ab} = \left( \begin{array}{cc}  i\partial_t + J_{ij}(t) & 0\\ 
0 &  -i\partial_t + J_{ij}^*(t) \end{array} \right)_{ab}.
\end{eqnarray}
We define the vacuum expectation value (VEV) $\varphi(x)$ and the quasiparticle (phonon) propagator $G(x,y)$ as
\begin{equation}
\varphi_a(x) = \langle b_a(x)\rangle,\ \  iG_{ab}(x,y) = \langle b_a(x) b_b^*(y)\rangle_c = \left(\begin{array}{cc} 
\langle \tilde b(x) \tilde b^*(y)\rangle_c & \langle \tilde b(x) \tilde b(y)\rangle_c \\
\langle \tilde b^*(x) \tilde b^*(y)\rangle_c & \langle \tilde b^*(x) \tilde b(y)\rangle_c
\end{array} \right).
\end{equation}
The microscopically occupied fields are denoted with a tilde $\tilde b(x)$. Hence the Green's function $G$ defined above does \emph{not} include the condensate fraction. The effective action is given by the double Legendre transform of the original action w.r.t.~the VEV $\varphi(x)$ and the correlator $G_{ab}(x,y)$\cite{cornwall_74,berges_04}:
\begin{eqnarray}
\Gamma[\varphi,G] &=& S[\varphi,\varphi^*] + \frac{1}{2}\text{Tr}[\log G^{-1}] + \frac{1}{2}\text{Tr}[G_0^{-1}(\varphi)G] - \Gamma_2[\varphi,G],\nonumber\\
S[\varphi,\varphi^*] &=& \int\mathrm{d}x\mathrm{d}y \varphi^*(x)G^{-1}_\text{free}(x,y)\varphi(y) - \frac{U}{2}\int\mathrm{d}x\left|\varphi(x)\right|^4 \equiv \int\mathrm{d}x\mathrm{d}y \varphi^*(x)G^{-1}_0(x,y;\varphi)\varphi(y),
\end{eqnarray}
where the sum over the Bogoliubov-Nambu index $a$ is implicit. The Bogoliubov propagator $G^{-1}_0(x,y;\varphi)$ generates the motion of the Gross-Pitaevskii equation. Notice that it depends on the field $\varphi$ itself since the GPE is nonlinear. From that we obtain the inverse Bogoliubov propagator $\left(G_0^{-1}\right)_{ab}(x,y;\varphi)$ via:
\begin{eqnarray}
\frac{1}{2}\left(G_0^{-1}\right)_{ab}(x,y;\varphi) &=& \frac{\delta^2 S[\varphi,\varphi^*]}{\delta \varphi_{a}^*(x)\delta \varphi_{b}(y)} = \frac{1}{2}\left(G_\text{free}^{-1}\right)_{ab}(x,y) - \frac{U}{2}\delta_C(x-y)\left(\begin{array}{cc}
2|\varphi(x)|^2 & \varphi(x)^2\\
(\varphi(x)^*)^2 & 2|\varphi(x)|^2
\end{array}\right)_{ab}.\nonumber\\
\end{eqnarray}
So far the calculation is exact, however, we have not specified the Luttinger-Ward functional $\Gamma_2[\varphi,G]$ yet which is the sum of all two-particle irreducible diagrams and thus has to be treated approximately. Here, we consider a weak-coupling expansion which amounts to consider diagrams to first order in $U$, see Fig.~\ref{fig:Gamma2_nonC}.

\begin{figure}
	\centering
	\includegraphics[width = 1\columnwidth]{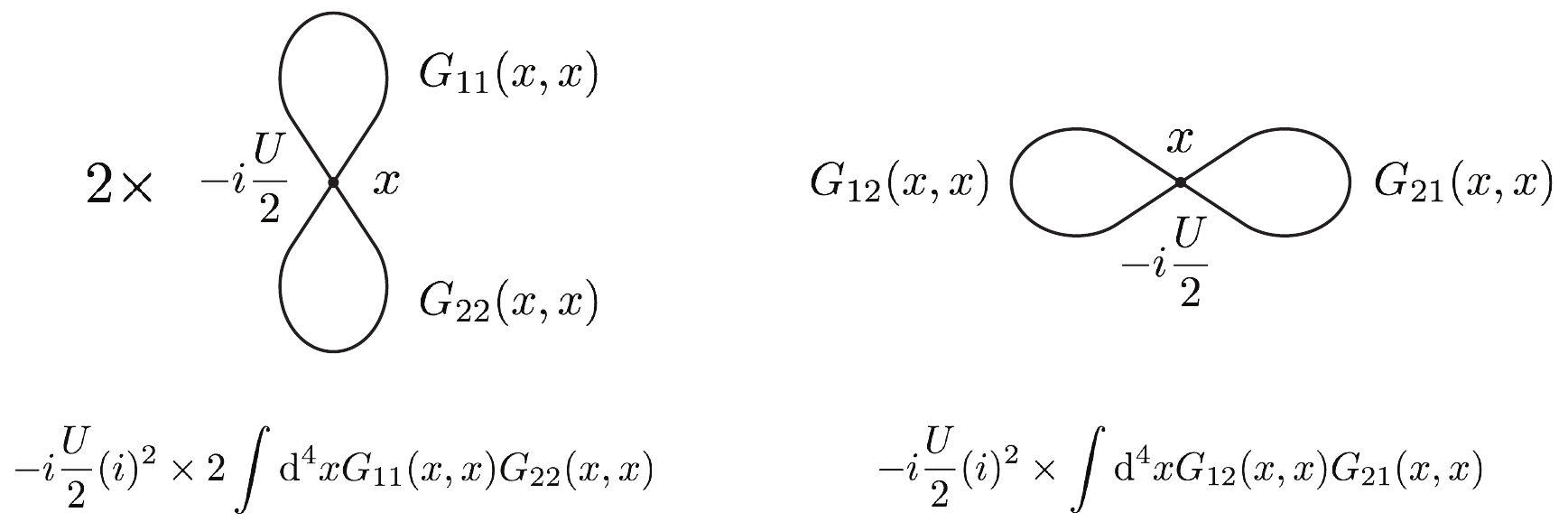}
	\caption{\label{fig:Gamma2_nonC} All two-particle irreducible diagrams which enter $\Gamma_2$ to first order in $U$, with their proper combinatorial factors. The diagrams can be turned into equations using the following Feynman rules: (i) a factor of $-iU/2$ for each vertex, and (ii) a factor of $i$ for each closed loop. By symmetry $G_{11} = G_{22}$ and $G_{12} = \left(G_{21}\right)^*$.}
\end{figure}

The EOM for the VEV and the propagator are obtained by making the effective action $\Gamma$ stationary with respect to the fields, $\frac{\delta\Gamma[\varphi,G]}{\delta\varphi^*}=0$ and $\frac{\delta\Gamma[\varphi,G]}{\delta G_{ab}}=0$, which lead to:
\begin{eqnarray}
\int_C\mathrm{d}y \left(G_\text{free}^{-1}\right)_{11}(x,y)\varphi(y) - U \varphi^*(x)\varphi^2(x)  -  U\left(  2\varphi(x) G_{11}(x,x) + \varphi^*(x)G_{12}(x,x) \right) &=& 0,\nonumber\\
\sum_b\left[\left(G_\text{free}^{-1}\right)_{ab}(t) - U\delta_C(x-y) \left(\begin{array}{cc}
2(iG_{11}+|\varphi|^2) &  iG_{12} + \varphi^2 \\
iG_{21} + (\varphi^*)^2  & 2(iG_{22} + |\varphi|^2)
\end{array}\right)_{ab} \right]G_{bc}(t,t') &=& \delta_{ac}\delta_C(t-t'),
\label{eq:EOM_phi_nonC}
\end{eqnarray} 
and the Green's function multiplication in the second equation above is understood in the matrix-multiplication sense: $(AB)(x,z) = \int_y A(x,y)B(y,z)$. We remark that these EOM are equivalent to the Bogoliubov-Hartree-Fock EOM derived in Ref.~\onlinecite{griffin_95} when starting from the lab frame Hamiltonian (see main text), making the ansatz $b_{k=0} = \varphi +\tilde b_{k=0}$, $b_{k\neq 0} =\tilde b_{k\neq 0}$, and then linearises any cubic terms in $\tilde b_k$.

Next, we open the closed time contour\cite{berges_04} by decomposing the Green's function into a spectral part $\rho(x,y)$ and a statistical part $F(x,y)$ according to
\begin{eqnarray}
iG(x,x') &=& F(x,x') - \frac{i}{2}\rho(x,x')\text{sgn}_C(t-t'),\nonumber\\
F_{ab}(x,x') &=& \frac{1}{2}\langle\{b_a(x),b_b^\dagger(x')\}\rangle_c = \frac{1}{2}\left(\begin{array}{cc}
\langle\{\tilde b(x),\tilde b^\dagger(x')\}\rangle & \langle\{\tilde b(x),\tilde b(x')\}\rangle \\
\langle\{\tilde b^\dagger(x),\tilde b^\dagger(x')\}\rangle & \langle\{\tilde b^\dagger(x),\tilde b(x')\}\rangle
\end{array}\right)_{ab}\nonumber\\
\rho_{ab}(x,x') &=& i\langle[b_a(x),b_b^\dagger(x')]\rangle_c = i\left(\begin{array}{cc}
\langle[\tilde b(x),\tilde b^\dagger(x')]\rangle & \langle[\tilde b(x),\tilde b(x')]\rangle \\
\langle[\tilde b^\dagger(x),\tilde b^\dagger(x')]\rangle & \langle[\tilde b^\dagger(x),\tilde b(x')]\rangle.
\end{array}\right)_{ab}
\end{eqnarray}
The following relations follow immediately from the above definitions:
\begin{eqnarray}
F_{12}(x,x') &=& F_{12}(x',x),\ \ \ \rho_{12}(x,x') = -\rho_{12}(x',x),\nonumber\\
F_{21}(x,x') &=& F_{21}(x',x),\ \ \ \rho_{21}(x,x') = -\rho_{21}(x',x),\nonumber\\
F_{12}(x,x') &=& F_{21}^*(x,x'),\ \ \ \rho_{12}(x,x') = \ \ \rho_{21}^*(x,x'),\nonumber\\
F_{11}(x,x') &=& F_{11}^*(x',x),\ \ \ \rho_{11}(x,x') = -\rho_{11}^*(x',x),\nonumber\\
F_{22}(x,x') &=& F_{11}^*(x,x'),\ \ \ \rho_{22}(x,x') =\ \ \rho_{11}^*(x,x').
\label{eq:symm_rel}
\end{eqnarray}
We now assume that the system is translationally invariant, with periodic boundary conditions. We find the following system of coupled nonlinear EOM in momentum space for the condensate
\begin{eqnarray}
i\partial_t\varphi(t) &=&  (zJ_0-\mu)\varphi(t) + \varepsilon_{k=0}(t)\varphi(t)\nonumber\\
&&  + \frac{U}{N_s}\left[ \left[\varphi(t)\right]^*\left[\varphi(t)\right]^2 + 2\varphi(t) \int_q F_{11}(t,t;q) + \left[\varphi(t)\right]^*\int_q F_{12}(t,t;q) \right],\nonumber\\
\label{eq:WCCA_GPE}
\end{eqnarray}
and the statistical correlator $F$
\begin{eqnarray}
i\partial_tF_{11}(t,t';k) &=& (zJ_0-\mu)F_{11}(t,t';k) + \varepsilon_k(t)F_{11}(t,t';k) \nonumber\\
&& + \frac{U}{N_s}\left[ 2\left( |\varphi(t)|^2 + \int_q F_{11}(t,t;q) \right)F_{11}(t,t';k) + \left(\left[\varphi(t)\right]^2 + \int_q F_{12}(t,t;q)\right) \left[F_{12}(t,t';k)\right]^*  \right],\nonumber\\
i\partial_tF_{12}(t,t';k) &=& (zJ_0-\mu)F_{12}(t,t';k) + \varepsilon_k(t)F_{12}(t,t';k)\nonumber\\
&&+ \frac{U}{N_s}\left[ 2\left( |\varphi(t)|^2 + \int_q F_{11}(t,t;q) \right)F_{12}(t,t';k) + \left( \left[\varphi(t)\right]^2 + \int_q F_{12}(t,t;q)\right)\left[F_{11}(t,t';k)\right]^*  \right].\nonumber\\
\end{eqnarray}
For completeness, we also give the equations of motion for the spectral correlators $\rho$ which, on the other hand, obey
\begin{eqnarray}
i\partial_t\rho_{11}(t,t';k) &=& (zJ_0-\mu)\rho_{11}(t,t';k) + \varepsilon_k(t)\rho_{11}(t,t';k) \nonumber\\
&& + \frac{U}{N_s}\left[ 2\left( |\varphi(t)|^2 + \int_q F_{11}(t,t;q) \right)\rho_{11}(t,t';k) + \left( \left[\varphi(t)\right]^2 + \int_q F_{12}(t,t;q)\right)\left[\rho_{12}(t,t';k)\right]^*  \right],\nonumber\\
i\partial_t\rho_{12}(t,t';k) &=& (zJ_0-\mu)\rho_{12}(t,t';k) + \varepsilon_k(t)\rho_{12}(t,t';k)\nonumber\\
&&+ \frac{U}{N_s}\left[ 2\left( |\varphi(t)|^2 + \int_q F_{11}(t,t;q) \right)\rho_{12}(t,t';k) + \left( \left[\varphi(t)\right]^2 + \int_q F_{12}(t,t;q)\right)\left[\rho_{11}(t,t';k)\right]^*  \right].\nonumber\\
\end{eqnarray}
In the above equations, $z$ is the coordination number, $\varepsilon_k(t)$ is the time-periodic free dispersion in the rotating frame, and the integrals are all taken over the Brillouin zone.

Furthermore, if one is interested in the equal-time correlation of the statistical correlator $F$, using the symmetry relations in Eq.~\eqref{eq:symm_rel} one arrives at the somewhat simplified equations
\begin{eqnarray}
\partial_t F_{11}(t,t;k) &=& 2\text{Im}\left\{  \frac{U}{N_s}\left( \left[\varphi(t)\right]^2 + \int_q F_{12}(t,t;q)\right)\left[F_{12}(t,t;k)\right]^* \right\},\nonumber\\
i\partial_tF_{12}(t,t;k) &=& 2\bigg\{ (zJ_0-\mu)F_{12}(t,t;k) + \varepsilon_k(t)F_{12}(t,t;k) \nonumber\\
&&+ \frac{U}{N_s}\left[ 2\left( |\varphi(t)|^2 + \int_q F_{11}(t,t;q) \right)F_{12}(t,t;k) + \left( \left[\varphi(t)\right]^2 + \int_q F_{12}(t,t;q)\right)\left[F_{11}(t,t;k)\right]^*  \right]\bigg\}.
\label{eq:WCCA_F}
\end{eqnarray}

\subsection{\label{app:AB} WCCA Equations of Motion for the BHM on a General Bipartite Lattice}

We now generalise the WCCA EOM to any bipartite lattice. Consider a bipartite lattice with the two sublattices labelled by $A$ and $B$ and periodic boundary conditions. Each sublattice contains $N_A = N_B = N_s/2$ number of sites. With this definition, additionally to the Bogoliubov index $a = 1,2$, all correlators carry an additional index $\alpha = A, B$, and so does the condensate fraction.

The extended system of equal-time equations of motion for the WCCA of the BHM reads
\begin{eqnarray}
i\partial_t\varphi^A(t) &=&  (zJ_0-\mu)\varphi^A(t) + \varepsilon_{k=0}(t)\varphi^B(t)\nonumber\\
&&  + \frac{U}{N_A}\left[ \left[\varphi^A(t)\right]^*\left[\varphi^A(t)\right]^2 + 2\varphi^A(t) \int_q F^{AA}_{11}(t,t;q) + \left[\varphi^A(t)\right]^*\int_q F_{12}^{AA}(t,t;q) \right],\nonumber\\
i\partial_t\varphi^B(t) &=&  (zJ_0-\mu)\varphi^B(t) + \varepsilon_{k=0}(t)\varphi^A(t)  \nonumber\\
&& + \frac{U}{N_A}\left[ \left[\varphi^B(t)\right]^*\left[\varphi^B(t)\right]^2  + 2\varphi^B(t)\int_q F^{BB}_{11}(t,t) + \left[\varphi^B(t)\right]^*\int_q F_{12}^{BB}(t,t) \right],
\label{eq:WCCA_GPE_AB}
\end{eqnarray}

\begin{allowdisplaybreaks}
	\begin{eqnarray}
	\partial_t F^{AA}_{11}(t,t;k) &=& 2\text{Im}\left\{ \varepsilon_k(t)\left[F^{AB}_{11}(t,t;k)\right]^* + \frac{U}{N_A}\left( \left[\varphi^A(t)\right]^2 + \int_q F^{AA}_{12}(t,t;q)\right)\left[F^{AA}_{12}(t,t;k)\right]^* \right\},\nonumber\\
	\partial_t F^{BB}_{11}(t,t;k) &=& 2\text{Im}\left\{ \varepsilon^*_k(t)F^{AB}_{11}(t,t;k) + \frac{U}{N_A}\left( \left[\varphi^B(t)\right]^2 + \int_q F^{BB}_{12}(t,t;q)\right)\left[F^{BB}_{12}(t,t;k)\right]^* \right\},\nonumber\\
	i\partial_t F^{AB}_{11}(t,t;k) &=& \varepsilon_k(t)\left( F^{BB}_{11}(t,t;k) - F^{AA}_{11}(t,t;k)  \right) + \frac{U}{N_A}\bigg[ \nonumber\\
	&& + 2\left( |\varphi^A(t)|^2 - |\varphi^B(t)|^2 + \int_q F^{AA}_{11}(t,t;q) - F^{BB}_{11}(t,t;q) \right)F^{AB}_{11}(t,t;k)\nonumber\\
	&& + \left( \left[\varphi^A(t)\right]^2 +  \int_q F^{AA}_{12}(t,t;q) \right)\left[F^{AB}_{12}(t,t;k)\right]^*
	- \left( \left[\varphi^B(t)\right]^2 +  \int_q F^{BB}_{12}(t,t;q) \right)^*F^{AB}_{12}(t,t;k) \bigg],\nonumber\\
	i\partial_tF^{AA}_{12}(t,t;k) &=& 2\bigg\{ (zJ_0-\mu)F^{AA}_{12}(t,t;k) + \varepsilon_k(t)F^{BA}_{12}(t,t;k) \nonumber\\
	&&+ \frac{U}{N_A}\left[ 2\left( |\varphi^A(t)|^2 + \int_q F^{AA}_{11}(t,t;q) \right)F^{AA}_{12}(t,t;k) + \left( \left[\varphi^A(t)\right]^2 + \int_q F^{AA}_{12}(t,t;q)\right)\left[F^{AA}_{11}(t,t;k)\right]^*  \right]\bigg\} ,\nonumber\\
	i\partial_tF^{BB}_{12}(t,t;k) &=& 2\bigg\{\varepsilon_k^*(t)F^{AB}_{12}(t,t;k) + (zJ_0-\mu)F^{BB}_{12}(t,t;k)  \nonumber\\
	&& + \frac{U}{N_A}\left[ 2\left( |\varphi^B(t)|^2 + \int_q F^{BB}_{11}(t,t;q)\right) F^{BB}_{12}(t,t;k) + \left( \left[\varphi^B(t)\right]^2 +  \int_q F^{BB}_{12}(t,t;q) \right)\left[F^{BB}_{11}(t,t;k)\right]^*  \right]\bigg\} ,\nonumber\\
	i\partial_tF^{AB}_{12}(t,t;k) &=& 2(zJ_0-\mu)F^{AB}_{12}(t,t;k) + \left(\varepsilon_k(t)^*F^{AA}_{12}(t,t;k) + \varepsilon_k(t)F^{BB}_{12}(t,t;k)\right) + \frac{U}{N_A}\bigg[ \nonumber\\
	&& + 2\left( |\varphi^A(t)|^2 + |\varphi^B(t)|^2 + \int_q F^{AA}_{11}(t,t;q) + F^{BB}_{11}(t,t;q) \right)F^{AB}_{12}(t,t;k) \nonumber\\
	&& + \left( \left[\varphi^A(t)\right]^2 +  \int_q F^{AA}_{12}(t,t;q) \right)\left[F^{AB}_{11}(t,t;k)\right]^*
	- \left( \left[\varphi^B(t)\right]^2 +  \int_q F^{BB}_{12}(t,t;q) \right)^*F^{AB}_{11}(t,t;k) \bigg].
	\label{eq:WCCA_FAB}
	\end{eqnarray}
\end{allowdisplaybreaks}
All integrals in Eqs.~\eqref{eq:WCCA_GPE_AB} and~\eqref{eq:WCCA_FAB} are taken over the reduced Brillouin zone (w.r.t.~the $AB$-sublattice symmetry). Equations~\eqref{eq:WCCA_GPE_AB} and~\eqref{eq:WCCA_FAB} constitute a coupled set of non-linear equations, the solution of which produces the dynamics discussed in the main text. Note that these EOM can be applied to systems with arbitrary time-dependence (not necessarily a periodic one) and on an arbitrary bipartite lattice, such as the honeycomb lattice.

For the analysis in the main text, the initial condition for the condensate fractions is chosen to be $|\varphi^{A}(0)|^2/N_s = n_0/2$ for $\zeta < 2.405$ and $|\varphi^{B}(0)|^2/N_s = n_0/2$ for $\zeta > 2.405$, where $n_0$ is the total condensate fraction for the non-driven model in Bogoliubov theory. 

\section{Validity of the WCCA and Thermalisation Timescales }

In this Appendix we estimate the timescales on which the WCCA gives a reliable description of the physics, and discuss the dominant processes that (in the weak coupling regime) destabilize the various pre-thermal steady states discussed in the main text. We discuss each of the three regimes separately.

\emph{Parametrically unstable region}. In this regime, the prethermalized phase is the one in which the momentum distribution is sharply peaked along momentum-space arcs as in Fig.~3 (a) (see main text). As in the main text, we treat the time-averaged dispersion as the unperturbed Hamiltonian and look at the parametric instability growth rate growth rate due to a perturbation of the form $J_0 \mathcal{J}_1(\zeta) b^\dagger_{\mathbf{k}} b_{\mathbf{k}}e^{i\Omega t}$. 
The matrix element for pair creation is then $\sim U n_0 J_0\mathcal{J}_1(\zeta)/W_\mathrm{ave}$ (the Fourier coefficient $c^{l}(\zeta)$ from the parametric instability analysis above is essentially given by the Bessel function), and for reasonably large drives this is linear in $U$.
Parametric instability predicts that these features will grow at the rate $\Gamma \sim U n_0$, where $n_0$ is the condensate amplitude. The decay rate (i.e., inverse lifetime) of the quasiparticles along these arcs, once they are formed, is limited by collisions, and Fermi's Golden Rule implies that this decay rate is of order $U^2$; this is the rate at which these features spread out in momentum space. Thus there is a	 parametric separation in $U$ between the formation and decay rate of these peaks. The leading collisional process comes from cubic terms of the form $U \varphi^* b^\dagger_{\mathbf{k}_1} b_{\mathbf{k}_2} b_{\mathbf{k}_1 - \mathbf{k}_2}$ (plus appropriate conjugates) in the Hamiltonian. The Golden Rule rate for this particular process is 

\begin{equation}
\Gamma_c(\mathbf{k})  \sim U^2 n_0 n_{\mathbf{k}} \mathcal{N}_{2p}(E_\text{ave}(\mathbf{k}))
\end{equation}
where $\mathcal{N}_{2p}(E_\text{ave}(\mathbf{k})) \sim \int \mathrm{d}E' d^2 q \delta(E' - E_\text{ave}(\mathbf{q})) \delta(E_\text{ave}(\mathbf{k}) - E' - E_\text{ave}(\mathbf{q - k}))$ is the accessible two-particle density of states. Here, $E_\text{ave}(\mathbf{q})$ is the energy of an excitation with quasimomentum $\mathbf{q}$. On dimensional grounds this two-particle density of states must be inversely proportional to $W_{\mathrm{ave}}$; thus, the overall Golden Rule lifetime of a particular quasiparticle state will go as

\begin{equation}
\Gamma_c(\mathbf{k}) \sim U^2 n_0 n_{\mathbf{k}} / W_\text{ave}
\end{equation}
up to a multiplicative constant. The ratio between the decay rate and the creation rate scales as $Un_{\bf k}/(J_0\mathcal{J}_1(\zeta))$. Thus the decay rate of a mode is slower than the creation rate whenever the condensate amplitude is large compared with the population of the mode (essentially because the matrix element is not Bose-enhanced to the same degree). However, the decay rate is also suppressed with decreasing the interactions (expected) or increasing the drive amplitude. 
At short times, when the condensate is not appreciably depleted, the WCCA is therefore reliable; however, when the depletion becomes large the WCCA also fails. Thus the regimes of validity of the WCCA and Bogoliubov theory in the parametrically unstable regime are essentially the same, although the WCCA has the advantage of respecting particle number conservation \emph{exactly} at all times.  

\emph{Stable region.} In the stable region there are two types of physical processes beyond the WCCA. (1) The excitations created by the original quench into the phase have finite collisional lifetimes, as discussed above. The momentum-space patterns in the stable region will dephase on this Golden-Rule timescale $\Gamma_c$; however, the condensate fraction will remain large and stable even after dephasing. (2) Eventually, the system will absorb energy from the drive. If the drive frequency is $\Omega$ and the bandwidth of single-particle excitations is $W_{\mathrm{ave}}$, then resonant absorption must involve at least $m \equiv \Omega / W_{\mathrm{ave}}$ quasiparticles. It is straightforward to check that the associated Golden Rule rate, at weak coupling, is of the form $U^m/W_{\mathrm{ave}}^{m - 1}$. When $U$ is sufficiently small, this heating timescale is much longer than the timescale on which the momentum-space patterns dephase; thus the system should remain \emph{stable} for extremely long times at 
high frequencies.

\emph{Dynamically unstable region.} In this regime, the growth rates of unstable modes are of order $\sqrt{J_\mathrm{ave}(\zeta)U}$, whereas the collision rates are of order $U^2/W_\text{ave}$ at best, so at weak coupling we have a parametric window in $U$ where the WCCA remains valid. 

\section{Phase Transition Region around $\zeta  =2.405$}

In this appendix, we discuss the dynamics governed by the WCCA close to the first zero of the Bessel function, $\zeta = 2.405$, where the dispersion of the $\Omega\to\infty$ Hamiltonian becomes flat (central grey region in Fig.~1, main text). For $\zeta < 2.405$ the dispersion of the free theory $U=0$ supports a stable minimum for ${\bf k} = {\bf 0}$, while for $\zeta > 2.405$ the stable minimum appears at ${\bf k} = \boldsymbol{\pi}$. Since the the two stable regions support different momentum modes, a phase transition occurs in between them. Therefore, it is required that one allows for a macroscopic population of both the modes in the immediate vicinity of $\zeta=2.405$. 

This can be achieved by reducing the translational symmetry of the problem. Intuitively, a condensate at ${\bf k} = {\boldsymbol{\pi}}$ with amplitude $\varphi_{k=\pi}$ flips a sign on every other site. Hence, one can choose to work in the original (momentum-resolved) basis $(\varphi_{k=0},\varphi_{k=\pi})$, or in the site-resolved basis $(\varphi^A,\varphi^B)$. The two are related by a rotation. In order allow for a dynamical population of the $\varphi_{k=\pi}$ condensate, Eqs.~\eqref{eq:WCCA_GPE_AB} and~\eqref{eq:WCCA_FAB} require that the initial condition for $\varphi_{\pi}(0) = 1/\sqrt{2}\left(\varphi^A(0) - \varphi^B(0)\right)$ be nonzero. In the $AB$-basis, this is equivalent to saying that there is a slight difference in the condensate occupation on the two sublattices. Physically, this imbalance is caused by spontaneous symmetry breaking. However, in the WCCA one has to put in this imbalance by hand. In the following we refer to the small value $s = |\varphi_{\pi}(0)|^2$ as \emph{seed}. 

\begin{figure}
	\centering
	\includegraphics[width = 0.4\columnwidth]{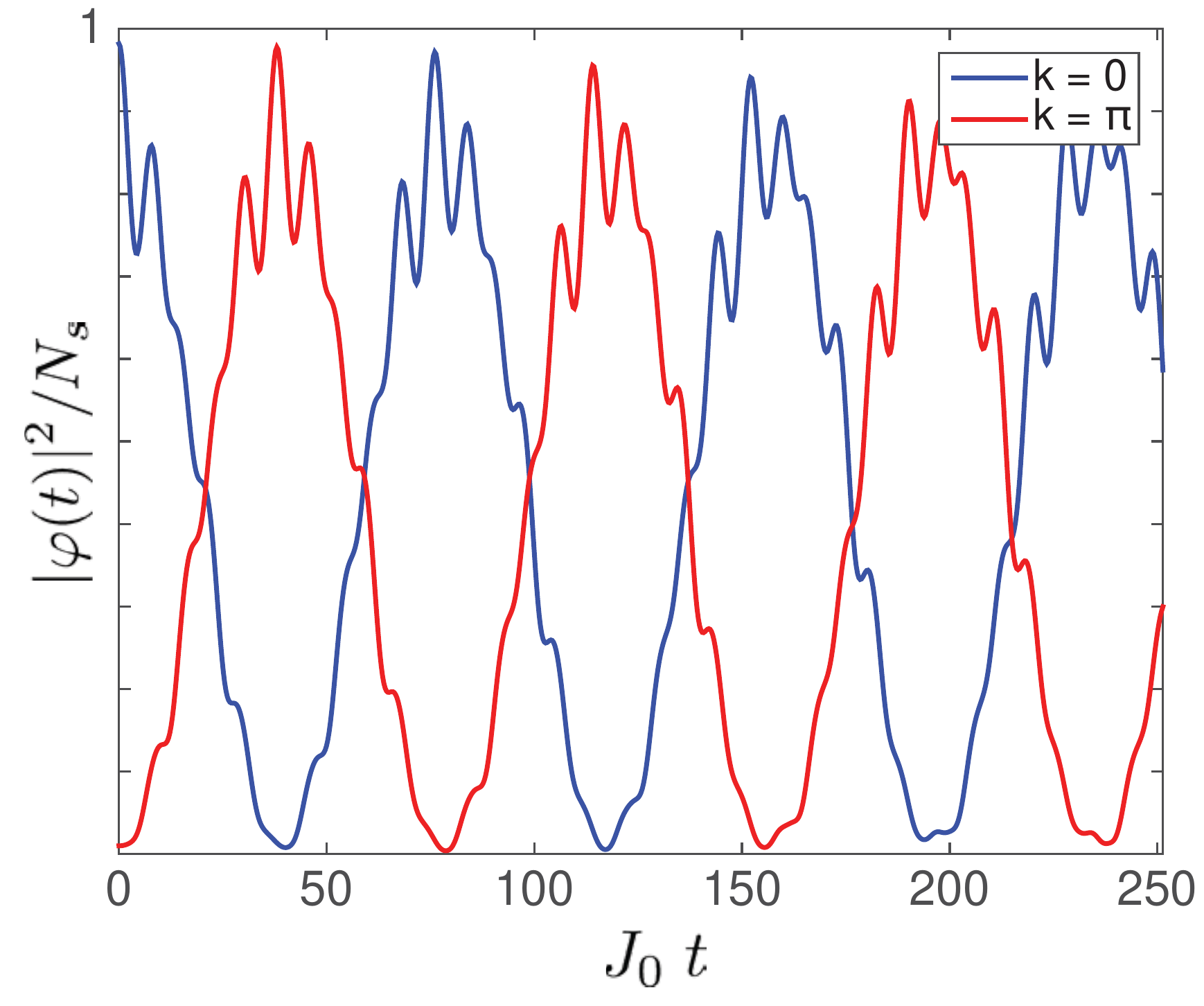}
	\caption{\label{fig:osc_condensate} Time evolution of the condensate fractions for $\zeta = 2.405$ and $\Omega/J_0 = 20$ starting from a Bogoliubov initial state localised at ${\bf k} = {\bf 0}$. The seed size is $s = 1\%$ and $U/J_0=0.2$.}
\end{figure}

When the effective dispersion becomes flat $\zeta \approx 2.405$ [Fig~\ref{fig:osc_condensate}], the condensate undergoes oscillations between the ${\bf 0}$ and $\boldsymbol{\pi}$ modes, with a period $\sim 1/U$ for small $U$. This behaviour is reminiscent of the collapse-and-revival effects seen for a BEC that is suddenly quenched into the Mott insulating phase\cite{greiner_02}, although the dynamics governed by WCCA is classical. The period of the transfer oscillations is also seed-dependent and increases with $s\to 0$. Even though our approximation does not capture a true Mott insulating phase, the nonlinearities included in the WCCA are sufficient to give rise to these oscillations. Physically speaking, a quasiparticle-mediated channel is opened, through which particles flow from the condensate at ${\bf k} = \mathbf{0}$ to ${\bf k} = \boldsymbol{\pi}$. Although it is present at any $\zeta$, this channel is only effective when the dispersion is sufficiently flat since the amplitude for the phonon-
mediated transition $\varphi_{k=0}\to b^\dagger_k \to \varphi_{k=\pi}$ scales as $\left(U/J_0\right)^2$.

\section{Time-Dependence of the Energy}

\begin{figure}[h!]
	\centering
	\includegraphics[width = 0.5\columnwidth]{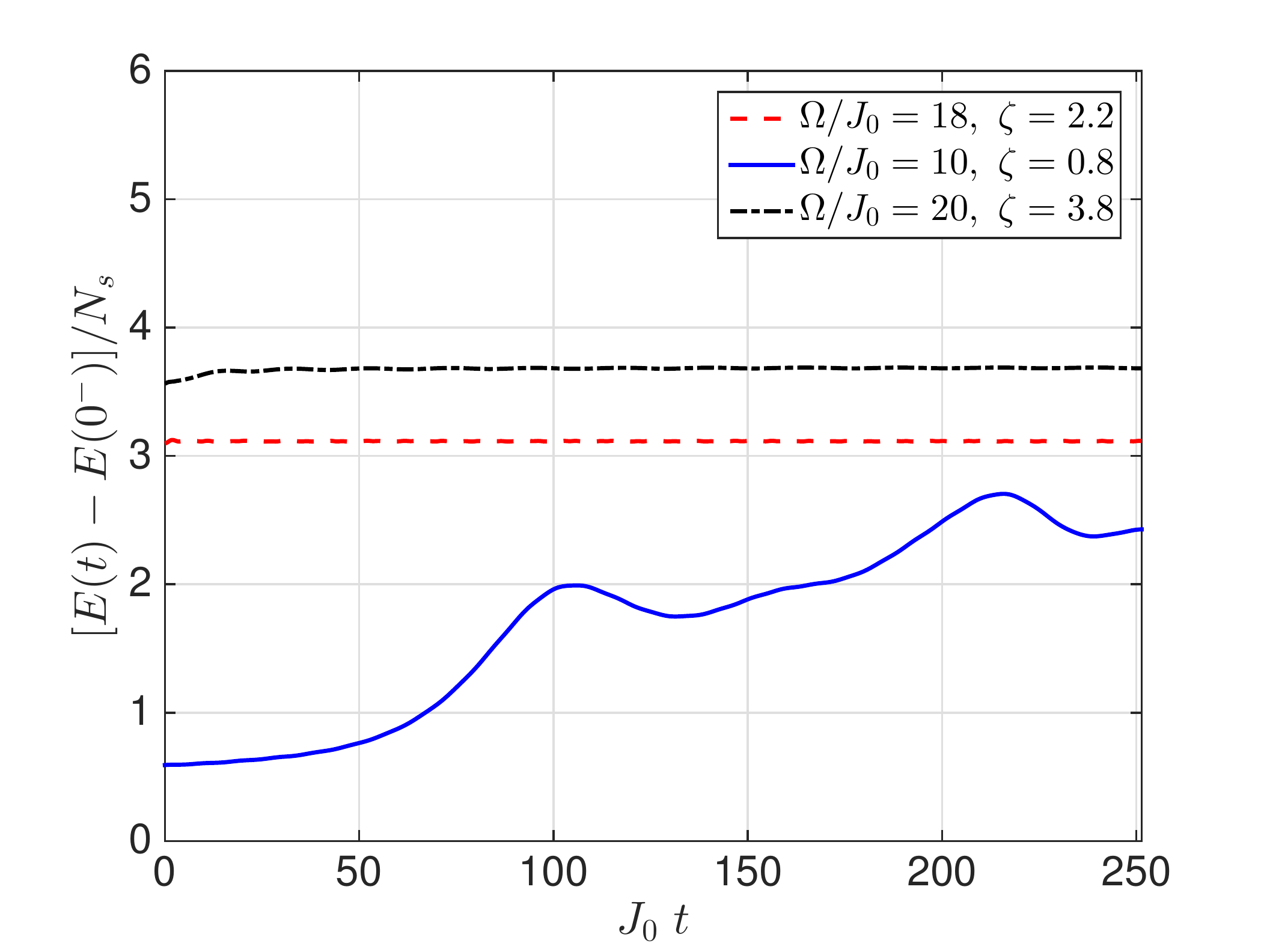}
	\caption{\label{fig:energy} Total energy density of condensate and quasiparticles as a function of time for $U/J_0=0.2$ following the quench with frequency and amplitude as stated in the legend.}  
\end{figure} 
Last, we briefly address the issue of heating. Fig.~\ref{fig:energy} shows the excess total (i.e., condensate plus quasiparticle) energy density in the system, relative to the non-driven state. Due to the abrupt turn-on of the periodic circularly polarised modulation, the energy changes discontinuously at $t=0$. 
As expected, the energy density increases due to heating in the parametrically unstable region, saturates in the stable region, and exhibits a small growth for $\zeta \approx 3.8$. Notice the different behaviour in the parametrically unstable region compared to the dynamically unstable one: while in the former the energy grows due to the population of modes lying on the high-energy surface, in the latter the dynamically unstable modes appear close enough to the origin [cf.~Fig.~3, panel (c) in main text] so that the growth in energy density past the quench value is not substantial.  
Note that the system does not heat up even at fairly long times whenever the parameters are chosen to be in the stable region of the stability diagram. 
Although ergodic periodically-driven systems are expected to eventually heat up to infinite temperature\cite{dalessio_13,dalessio_14,ponte_15,abanin_15}, in the weak-coupling limit this heating timescale (which is due to collisions between quasiparticles) is parametrically slower in the ``stable'' regimes of our phase diagram than in the ``unstable'' regimes. 
Thus, for a range of present-day experiments, we expect that in the stable high-frequency regime there is no significant heating on experimentally relevant timescales.

\section{Time-Dependence of the Momentum Distribution Function}

For the time-evolution of the momentum distribution function, we refer to the three videos in the supplementary material. The lower left panel shows the quasiparticle momentum distribution over the first Brillouin zone, while the upper left panel is a top view of the same. The upper right panel displays the time-evolution of the condensate fraction, while the lower right panel shows the energy density. The parameters for each simulation can be found in the title. The three movies correspond to the points in the stability diagram marked by (a), (b), (c) in Fig.~1 (main text).

\end{widetext}

\end{document}